\documentstyle[12pt]{article}

\newcommand{\be}{\begin{equation}}
\newcommand{\ee}{\end{equation}}
\newcommand{\bdm}{\begin{displaymath}}
\newcommand{\edm}{\end{displaymath}}
\newcommand{\bea}{\begin{eqnarray}}
\newcommand{\eea}{\end{eqnarray}}

\newcommand{\fs}{\; \; .}
\newcommand{\co}{\; \; ,}

\newcommand{\eff}{{e\hspace{-0.1em}f\hspace{-0.18em}f}}

\newcommand{\QCD}{\mbox{\scriptsize Q\hspace{-0.1em}CD}}
\newcommand{\indR}{\mbox{\scriptsize R}}
\newcommand{\indL}{\mbox{\scriptsize L}}

\newcommand{\lvac}{\langle 0|\,}
\newcommand{\rvac}{\,|0\rangle}

\newcommand{\Hbar}{\,\overline{\rule[0.7em]{0.6em}{0em}}\hspace{-0.8em}H}
\newcommand{\ubar}{\overline{\rule[0.42em]{0.4em}{0em}}\hspace{-0.5em}u}
\newcommand{\dbar}{\,\overline{\rule[0.65em]{0.4em}{0em}}\hspace{-0.6em}d}
\newcommand{\sbar}{\overline{\rule[0.42em]{0.4em}{0em}}\hspace{-0.5em}s}
\newcommand{\cbar}{\overline{\rule[0.42em]{0.4em}{0em}}\hspace{-0.5em}c}

\newcommand{\lbar}{\overline{\rule[0.75em]{0.3em}{0em}}\hspace{-0.4em}l}

\begin{document}
\begin{titlepage}
\begin{flushright}BUTP-94/18\end{flushright}
\rule{0em}{2em}\vspace{2em}
\begin{center}
{\LARGE {\bf Foundations and scope of chiral perturbation theory}}\\
\vspace{2em}
H. Leutwyler\\Institut f\"{u}r theoretische Physik der Universit\"{a}t
Bern\\Sidlerstr. 5, CH-3012 Bern, Switzerland\\
\vspace{3em}
{\bf Abstract} \\
\vspace{2em}
\parbox{30em}{
The aim of this introductory lecture is to review the arguments, according to
which the symmetry properties of the strong interaction reveal themselves at
low energies. I first discuss the symmetries of QCD, then sketch the method
used to work out their implications and finally take up a few specific issues,
where new experimental results are of particular interest to test the
predictions.}\\

\vspace{2em}
Talk given at the Workshop\\ "Chiral Dynamics: Theory and Experiment",
July 1994, MIT
\\ \vspace{10em} \rule{30em}{.02em}\\ {\footnotesize Work
supported in part by Schweizerischer Nationalfonds}
\end{center}
\end{titlepage}

Chromodynamics is a gauge theory. The form
of the interaction among the gluons and quarks is fully determined by gauge
invariance. This implies, in particular, that the various different
quark flavours, $u,d,\ldots$ interact with the gluons in precisely the same
manner. As far as the strong interaction is
concerned, the only distinction
between, say, an $s$-quark and a $c$-quark is that the mass is different.
In this respect, the situation is the same as in
electrodynamics, where the interaction of the charged leptons with the
photon is also universal, such that the only difference between
$e,\mu$ and $\tau$ is the mass.
As an immediate consequence, the properties of a bound state like the
$\Lambda_s=(uds)$ are identical with those of the $\Lambda_c=(udc)$,
except for the fact that $m_c$ is larger than $m_s$.

\def\Large{\large}
\section{Isospin symmetry}
A striking property of the observed pattern of bound states is that they
come in nearly degenerate {\it isospin} multiplets: $(p,n),$
$(\pi^+,\pi^0,\pi^-),
\;\ldots\;$
In fact, the splittings within these multiplets are so small that, for a long
time, isospin was taken for an
{\it exact} symmetry of the strong interaction; the observed
small
mass difference between neutron and proton or $K^0$ and $K^+$ was blamed on the
electromagnetic interaction.
We now know that this picture is incorrect: the
bulk of isospin breaking does not originate in the electromagnetic
fields, which surround the various particles, but is due to the fact that
the $d$-quark is somewhat heavier than the $u$-quark.

{}From a theoretical point of view, the quark masses are free
parameters ---
QCD makes sense for any value of $m_u,m_d,\ldots\;$ It
is perfectly
legitimate to compare the real situation with a theoretical one, where some of
the quark masses are given values, which differ from those found in
nature. In connection with isospin symmetry, the theoretical limiting case
of interest is a fictitious world, with
$m_u=m_d$. In this limit, the flavours
$u$ and $d$ become indistinguishable. The
Hamiltonian acquires an exact symmetry with respect to the transformation
\bdm \begin{array}{lll}u&\rightarrow&\alpha u +\beta d\\
d&\rightarrow&\gamma u + \delta d\end{array}\co\hspace{5em}
V=\left(\!\begin{array}{ll}\alpha&\!\!\beta\\
\gamma&\!\!\delta\end{array}\!\right)\co\edm
provided the $2\times 2$ matrix $V$ is unitary, $V\in\mbox{U(2)}$. Even for
$m_u\neq m_d$, the Hamiltonian of QCD is invariant under
a change of phase of the quark fields. The extra
symmetry, occurring if the masses of $u$ and $d$ are taken to be the same, is
contained in the subgroup SU(2), which results if the phase of the
matrix $V$ is subject to
the condition $\det V\!=\!1$. The above transformation law states that $u$ and
$d$ form an isospin doublet, while the remaining
flavours $s,\,c,\,\ldots\,$ are singlets.

In reality, $m_u$ differs from $m_d$. The isospin group SU(2) only represents
an {\it approximate} symmetry. The piece
of the QCD Hamiltonian, which breaks isospin symmetry, may be exhibited by
rewriting the mass term of the $u$ and $d$ quarks in the form
\bdm m_u\,\ubar u+m_d\,\dbar d=\mbox{$\frac{1}{2}$}(m_u+m_d)(\ubar u+
\dbar d)+\mbox{$\frac{1}{2}$}(m_d-m_u)(\dbar d-
\ubar u)\fs\edm
The remainder of the Hamiltonian is invariant
under isospin transformations and the same is true of the operator
$\ubar u+\dbar d$. The
QCD Hamiltonian thus consists of an isospin invariant part
$\Hbar_0$ and a
symmetry breaking term $\Hbar_{\mbox{\scriptsize sb}}$, proportional to
the mass difference $m_d-m_u$, \be \label{iso1}
H_{\QCD}=\,\Hbar_0
+\,\Hbar_{\mbox{\scriptsize sb}}\co\hspace{3em}
\Hbar_{\mbox{\scriptsize sb}}\,=\mbox{$\frac{1}{2}$}(m_d - m_u)
\int\!\!d^3\!x\, (\dbar d - \ubar u)\fs\ee
The strength of isospin breaking is controlled by the quantity
$m_d-m_u$, which plays the role of a {\it symmetry breaking
parameter}. The fact that the multiplets are nearly degenerate
implies that the operator $\Hbar_{\mbox{\scriptsize sb}}$
only represents a small
perturbation --- the mass difference $m_d-m_u$ must be very small.
QCD thus provides a remarkably simple explanation for the fact that the strong
interaction is nearly invariant under
isospin rotations: it so happens that the difference between $m_u$ and
$m_d$
is small --- this is all there is to it.

The symmetry breaking also shows up in the properties of the vector currents,
e.g. in those of $\ubar \gamma^\mu d$. The integral of the corresponding charge
density over space, $I^+=\int\!d^3\!x\,u^\dagger d$, is the isospin raising
operator, converting a $d$-quark into a $u$-quark. The divergence of the
current is given by
\be\label{iso2} \partial_\mu (\ubar \gamma^\mu d)
=i\,(m_u-m_d)\,\ubar d\co \ee
and only vanishes for $m_u=m_d$, the condition for the charge $I^+$ to
be conserved. In the symmetry limit, there are three such conserved charges,
the three components of isospin,
$\;\vec{\rule{0em}{0.65em}}\hspace{-0.25em}I=(I^1,I^2,I^3)$. The
isospin raising operator considered above is the combination $I^+=I^1+i\,I^2$.
Since $\Hbar_0$ is invariant under isospin rotations, it conserves isospin,
\be\label{iso3}
[\;\,\vec{\rule{0em}{0.65em}}\hspace{-0.25em}I\,,\,\Hbar_0\,]=0\fs\ee
%%%%%%%%%%%%%%%%%%%%%%%%%%%%%%%%%%%%%%%%%%%%%%%%%%%%%%%%%%%%%%%%%%%
\section{Eightfold way}
\label{ew}
On the basis of the few strange particles, which had been discovered in the
course of the 1950's,
Gell-Mann and Ne'eman \cite{Gell-Mann} inferred that the strong
interaction possesses a further approximate symmetry,
of the same qualitative nature as
isospin, but more strongly broken. The symmetry,
termed the {\it eightfold way}, played a decisive role in unravelling
the quark degrees of freedom. By now, it has become evident that the
mesonic and baryonic levels are indeed grouped
in multiplets of SU(3) --- singlets, octets, decuplets --- and there is also
good phenomenological support for the corresponding symmetry relations among
the various observable quantities.

In the framework of QCD, eightfold way symmetry occurs
in the theoretical limit, where the three lightest quarks are
given the same mass, $m_u=m_d=m_s$. The Hamiltonian then becomes invariant
under the transformation
\bdm \left(\!\begin{array}{c} u\\d\\
s \end{array}\!\right) \rightarrow
V\left(\!\begin{array}{c} u\\d\\s \end{array}\!\right)\hspace{3em}V\in
\mbox{SU(3)}\edm
of the quarks fields and the
spectrum of the theory consists of degenerate multiplets of this group. The
degeneracy is lifted by the mass differences
$m_s-m_d$ and $m_d-m_u$, which represent the symmetry breaking parameters in
this case.
Since the eightfold way does represent an approximate symmetry of the strong
interaction, both of these mass differences must be small. Moreover, the
observed level pattern requires
$|m_d-m_u|\,\ll\,|m_s-m_d|$.

Formally, the above discussion may be extended to include additional
fla\-vours. One may even consider
the theoretical limit, where all of the quarks are given the same
mass.
The extension, however, does not correspond to an
approximate symmetry. The lightest
pseudoscalar bound state with the quantum numbers of $\dbar c$, e.g.,
sits at $M_{D^+}\simeq 1.87\,\mbox{GeV}$. If the mass of the charmed quark is
set equal to $m_u$, this state becomes degenerate with the $\pi^+$.
Clearly, the mass
difference $m_c-m_u$, which
plays the role of a symmetry breaking parameter in this case, does not
represent a small perturbation. We do not know why the quark masses
follow the pattern observed in nature, nor do we understand the
equally queer pattern of lepton masses. It so happens that the mass
differences between $u,\,d$ and $s$ are small, such that the
eightfold
way represents a decent approximate symmetry.
%%%%%%%%%%%%%%%%%%%%%%%%%%%%%%%%%%%%%%%%%%%%%%%%%%%%%%%%%%%%%%%%
\section{Chiral symmetry}
\label{chir}
The approximate symmetries discussed above explain
why the bound states of QCD exhibit a multiplet pattern,
but they do not account
for an observation which is equally striking and which plays a crucial role in
strong interaction physics --- the mass gap of the theory, $M_\pi,\,$ is
remarkably small.
The approximate symmetry, hiding behind this observation, was discovered by
Nambu \cite{Nambu}. It originates in a phenomenon, which is well-known
from neutrino physics: right- and left-handed components of {\it
massless} fermions do not communicate.

The symmetry, which forbids right-left-transitions,
manifests itself in the properties of the axial vector currents, such as
$\ubar \gamma^\mu\gamma_5d$. The corresponding continuity equation reads
\be\label{chir1}  \partial_\mu(\ubar \gamma^\mu\gamma_5d)=i\,(m_u+m_d)\,
\ubar\gamma_5d\fs\ee
While the divergence of the vector current
$\ubar \gamma^\mu d$ is proportional to the difference $m_u-m_d$, the
one of the axial current is proportional to
the sum $m_u+m_d$. If the two masses are
set equal, the vector current is conserved and the
Hamitonian becomes symmetric with respect to isospin rotations. If they are
not only taken equal, but equal to zero, then the axial current is
conserved, too, such that the corresponding charge
$I_5^+=\int\!d^3\!x\,d^\dagger\gamma_5 u$ also commutes with the
Hamiltonian --- QCD acquires an additional symmetry.

The isospin operator $I^+$ converts a $d$-quark
into a $u$-quark, irrespective of the helicity. The operator $I_5^+$,
however, acts differently on
the right- and left-handed components. The sum $\frac{1}{2}(I^++I_5^+)$
takes a righthanded $d$-quark into a righthanded $u$-quark, but leaves
left-handed ones alone.
This implies that, for massless quarks, the
Hamiltonian is invariant with
respect to a set of {\it chiral}
transformations: independent
isospin rotations of the right- and left-handed components of $u$ and $d$,
\bdm\left(\!\!\begin{array}{c} u_{\indR}\\d_{\indR}
\end{array}\!\!\right)\rightarrow V_{\indR}
\left(\!\!\begin{array}{c} u_{\indR}\\d_{\indR}
\end{array}\!\!\right)\co\;\;\;\;
\left(\!\!\begin{array}{c} u_{\indL}\\d_{\indL}
\end{array}\!\!\right)\rightarrow V_{\indL}
\left(\!\!\begin{array}{c} u_{\indL}\\d_{\indL}
\end{array}\!\!\right)\co\;\;\;\;\;\;\;V_{\indR},\,
V_{\indL} \in \mbox{SU(2)}\fs\edm
The corresponding symmetry group is the direct product of two separate isospin
groups, SU(2)$_{\indR}\times$SU(2)$_{\indL}$.
The symmetry is
generated two sets of isospin operators: ordinary isospin,
$\,\vec{\rule{0em}{0.65em}}\hspace{-0.25em}I$ and
chiral isospin, $\,\vec{\rule{0em}{0.65em}}\hspace{-0.25em}I_5$. The
particular operator considered above is
the linear combination $I_5^+=I^{\,1}_5+i\,I^{\,2}_5$.

In reality,
chiral symmetry is broken, because $m_u$ and $m_d$ do not vanish.
As above, the Hamiltonian may be split into a
piece which is invariant under the symmetry group of interest and a piece
which breaks the symmetry. In the present case, the symmetry breaking part
is the full mass term of the $u$ and $d$ quarks,
\be\label{chir2} H_{\QCD}=H^{\prime}_0+
H^{\prime}_{\mbox{\scriptsize sb}}\;\;,\;\;
H^{\prime}_{\mbox{\scriptsize sb}}
=\int\!d^3\!x(m_u\,\ubar u+m_d\,\dbar d)\fs\ee
The symmetric part conserves ordinary as well as chiral isospin,
\be\label{chir3} [\;\,\vec{\rule{0em}{0.65em}}\hspace{-0.25em}I \,
,\,H^{\prime}_0\,]=0\co\hspace{3em}
 [\;\,\vec{\rule{0em}{0.65em}}\hspace{-0.25em}I_5
,\,H^{\prime}_0\,]=0\fs\ee
Note that the symmetry group exclusively
acts on $u$ and $d$ --- the
remaining quarks $s,c,\ldots$ are singlets. The corresponding mass terms
$m_s\sbar s+m_c\,\cbar c+\ldots$ do not break the symmetry and are included in
$H^{\prime}_0$.
%%%%%%%%%%%%%%%%%%%%%%%%%%%%%%%%%%%%%%%%%%%%%%%%%%%%%%%%%%%%%%%%%%%%%%%%%%
\section{Spontaneous symmetry breakdown}
\label{sp}
Much before QCD was discovered, Nambu pointed out that chiral symmetry
breaks down spontaneously.
The phenomenon plays a crucial role for the properties
of the strong interaction at low energy. To discuss it, I return to
the theoretical scenario,
where $m_u$ and $m_d$ are set equal to zero.

In this framework, isospin is conserved. The isospin group SU(2)
represents the prototype of a "manifest"
symmetry, with all the consequences known from quantum mechanics: (i) The
energy
levels form degenerate multiplets. (ii) The operators
$\,\vec{\rule{0em}{0.65em}}\hspace{-0.25em}I$ generate transitions within the
multiplets, taking a neutron, e.g., into a proton,
$I^+|n\rangle=|p\rangle$. (iii) The ground state is an isospin
singlet,
\be\label{sp1} \vec{\rule{0em}{0.65em}}\hspace{-0.25em}I\rvac=0\fs\ee

If chiral symmetry was realized in the same manner, the energy levels would be
grouped into degenerate multiplets of
the group SU(2)$_{\indR}\times$SU(2)$_{\indL}$. Since the chiral isospin
operators $\,\vec{\rule{0em}{0.65em}}\hspace{-0.25em}I_5$
carry negative parity, the multiplets would then necessarily contain members of
opposite parity. The listings of the Particle Data Group, however, do not show
any trace of such a pattern. A particle with the quantum
numbers of $I_5^+|n\rangle$ and nearly the same mass as the neutron, e.g.,
is not observed in nature.

In fact, the symmetry of the Hamiltonian does not ensure that the
corresponding
eigenstates form multiplets of the symmetry group. In particular, the
state with the lowest eigenvalue of the Hamiltonian need not be a
singlet. In the case of a magnet, e.g., the Hamiltonian is
invariant under rotations of the spin directions, but the ground state fails
to be invariant, because
the spins are aligned and thereby single out a direction. Whenever the state
with the lowest eigenvalue is less symmetric than the Hamiltonian, the symmetry
is called "spontaneously broken" or "hidden". Chiral symmetry belongs to
this category. For dynamical reasons, the most important state --- the
vacuum --- is symmetric only under ordinary isospin rotations, but does not
remain invariant if a chiral rotation is applied,
\be\label{sp2}\vec{\rule{0em}{0.65em}}\hspace{-0.25em}I_5\rvac\neq 0\fs\ee
Since the Hamiltonian commutes with chiral isospin, the three
states
$\,\vec{\rule{0em}{0.65em}}\hspace{-0.25em}I_5\rvac$ have the same energy
as the vacuum, $E=0$. The operators
$\,\vec{\rule{0em}{0.65em}}\hspace{-0.25em}I_5$ do
not carry momentum, either, so that the states
$\,\vec{\rule{0em}{0.65em}}\hspace{-0.25em}I_5\rvac$ have $\vec{P}=0$.
This indicates that the
spectrum of physical states contains three
massless particles. Indeed, the Goldstone theorem \cite{Goldstone}
rigorously shows that spontaneous symmetry breakdown gives rise to
massless particles, "Goldstone
bosons". Their quantum numbers are those of the states
$\,\vec{\rule{0em}{0.65em}}\hspace{-0.25em}I_5\rvac$: spin zero, negative
parity and $I=1$.

The three lightest mesons, $\pi^+\!,\pi^0\!,\pi^-$, carry precisely these
quantum numbers. The chiral isospin operators act like creation or
annihilation operators for pions: Applied to the vacuum, they generate a
state containing a pion, $I_5^+\rvac=|\pi^+\rangle$. Applied to a
neutron, they do not lead to a parity partner, but instead yield a state
containing a neutron and a pion, $I_5^+|n\rangle=|n\pi^+\rangle$, etc.
%%%%%%%%%%%%%%%%%%%%%%%%%%%%%%%%%%%%%%%%%%%%%%%%%%%%%%%%%%%%%%%%%%%%
\section{Pion mass}
\label{pm}
The above discussion concerns the theoretical world, where $u$ and $d$ are
assumed to be massless, such that the group
SU(2)$_{\indR}\times$SU(2)$_{\indL}$ represents an exact symmetry.
The Hamiltonian of QCD contains a quark mass term,
which breaks chiral symmetry.
To see how this affects the mass of the
Goldstone bosons,
consider the transition matrix element of the axial current
$\ubar\gamma^\mu\gamma_5d$,
from the vacuum to a one-pion state. Lorentz invariance
implies that this matrix element is determined by the pion
momentum $p^\mu$, up to a constant,
\bdm\langle\pi^+(p)|\,\ubar(x)\gamma^\mu\gamma_5d(x)\rvac
=-ip^\mu\sqrt{2}\,F_\pi\, e^{ipx}\fs\edm
The value of the constant is measured in pion decay, $F_\pi\simeq
93\;\mbox{MeV}$. For the divergence
$\partial_\mu(\ubar\gamma^\mu\gamma_5d)$, this yields an expression
proportional to
$p^2=M_{\hspace{-0.07em}\pi^{\hspace{-0.07em}+}}^2$. Denoting the analogous
matrix element of the pseudoscalar density by $G_\pi$,
\bdm\langle\pi^+(p)|\,\ubar(x)\gamma_5d(x)\rvac
=\; i\sqrt{2}\,G_\pi\, e^{ipx}\co\edm
the conservation law (\ref{chir1}) thus implies the exact
relation
\be\label{pm1}
M_{\hspace{-0.07em}\pi^{\hspace{-0.07em}+}}^2=(m_u+m_d)\,(G_\pi/ F_\pi)\fs\ee
The relation confirms that, when the symmetry breaking parameters $m_u,m_d$ are
put equal to zero, the pion mass vanishes, independently of the
masses of the other quark flavours.
The group SU(2)$_{\indR}\times$SU(2)$_{\indL}$ then represents a
spontaneously broken, {\em exact} symmetry, with three strictly massless
Goldstone bosons. When the quark
masses are turned on, the Goldstone bosons pick up mass:
$M_{\hspace{-0.07em}\pi^{\hspace{-0.07em}+}}$ grows in
proportion to
\raisebox{0.25em}{$\sqrt{\rule{4em}{0em}}$}$\hspace{-4em}m_u+m_d\,$.
The pions remain light, provided $m_u$ and
$m_d$ are small. The quark mass term of the Hamiltonian then
amounts to a small perturbation, such that the
group SU(2)$_{\indR}\times$SU(2)$_{\indL}$ still represents
an {\em approximate} symmetry, with
approximately massless
Goldstone bosons.

Moreover, as noted in section \ref{ew}, the observed level pattern also
requires the differences between $m_u,m_d$ and $m_s$ to be small.
Hence the strange quark must be light, too, such that the corresponding
mass term may also be treated as a perturbation. The
decomposition of the Hamiltonian then takes the form
\be\label{pm2} H_{\QCD}=H_0+
H_{\mbox{\scriptsize sb}}\;\;,\;\;
H_{\mbox{\scriptsize sb}}
=\int\!d^3\!x(m_u\,\ubar u+m_d\,\dbar d+m_s\,\sbar s)\fs\ee
The first term, $\mbox{H}_0$,
describes three massless flavours $(u,\,d,\,s)$ as well
as three massive ones $(c,\,b,\,t)$. It is symmetric with respect to
independent rotations of the
right- and left-handed components of $u,d$ and $s$, i.e., with respect to
the group $\mbox{SU(3)}_{\indR}\times
\mbox{SU(3)}_{\indL}$. The perturbation series, which results if
$H_{\mbox{\scriptsize sb}}$ is treated as a perturbation, amounts to an
expansion of the matrix elements and eigenvalues in powers of $m_u,m_d$
and $m_s$. The inequality
$|m_d-m_u|\ll |m_s-m_d|$, which
follows from the fact that isospin breaking is much smaller than the breaking
of eightfold way symmetry, implies that the $s$-quark is considerably
heavier than the other two, $m_u,\,m_d\ll m_s$.

The above arguments rely on two
phenomenological observations:

(a)$\;$
The pion mass is small compared to the masses of all other hadrons.
This indicates that the strong interaction possess an approximate,
spontaneously broken symmetry, with the pions as the corresponding
Goldstone bosons. Indeed, the Hamiltonian of QCD exhibits an approximate
symmetry with the proper quantum numbers, provided both $m_u$ and $m_d$
are small.

(b)$\;$
The multiplet structure seen in the particle data tables
indicates that the eightfold way is an approximate symmetry of the
strong interaction. For QCD to possess such a symmetry, the
mass differences $m_d-m_u$ and $m_s-m_d$ must be small.

Combining the two observations, one concludes
that the mass of the strange quark also amounts to a small perturbation:
The two groups SU(3) and
$\mbox{SU(2)}_{\indR}\times\mbox{SU(2)}_{\indL}$
can be approximate symmetries of the
QCD Hamiltonian only if
$\mbox{SU(3)}_{\indR}\times\mbox{SU(3)}_{\indL}$
represents an approximate symmetry, too.
The masses of the other quarks occurring in the Standard Model, on the other
hand, cannot be treated as a perturbation. Since the corresponding fields
$c(x),\,b(x)$ and $t(x)$ are singlets with respect to
SU(3)$_{\indR}\times$SU(3)$_{\indL}$, their contribution may be included in the
symmetric part of
the Hamiltonian, $H_0$. Their presence does not significantly affect the
low energy structure of the theory.

The decomposition of the QCD Hamiltonian in eq. (\ref{pm2}) may
be compared with the standard perturbative splitting
\bdm H_{\QCD}=H_{\mbox{\scriptsize free}}+ H_{\mbox{\scriptsize int}}\co\edm
where the first term describes free quarks and gluons, while the second
accounts for their interaction. The corresponding expansion parameter is
the coupling constant $g$. Since QCD
is asymptotically free, the effective coupling becomes weak
at large momentum transfers --- processes which exclusively involve large
momenta may indeed be analyzed by treating the interaction as a
perturbation. Perturbation theory, however, fails in
the low energy domain,
where the effective coupling is strong, such that it is not meaningful to
truncate the expansion in powers of $H_{\mbox{\scriptsize int}}$ after the
first few
terms. In particular, the structure of the ground state cannot be analyzed in
this way, while the above decomposition, which retains the
interaction
among the quarks and gluons in the "unperturbed" Hamiltonian $H_0$ and
only treats $m_u,\,m_d$ and $m_s$
as perturbations, is perfectly suitable for that purpose.
Note that the character of the
perturbation series
in powers of $H_{\mbox{\scriptsize sb}}$ is quite different from the
one in powers of $H_{\mbox{\scriptsize int}}$: while the eigenstates of
$H_{\mbox{\scriptsize free}}$ are
known explicitly, this is not the case with $H_0$, which still describes
a highly nontrivial, interacting system. $H_0$ differs from the full
Hamiltonian only in one respect: it possesses an exact group of chiral
symmetries.
%%%%%%%%%%%%%%%%%%%%%%%%%%%%%%%%%%%%%%%%%%%%%%%%%%%%%%%%%%%%%%%%%%%%%%%
\section{Quark masses}
\label{qm}
There is an immediate experimental check of the above theoretical arguments:
the spontaneous breakdown of the symmetry
$\mbox{SU(3)}_{\indR}\times\mbox{SU(3)}_{\indL}$ to
the subgroup $\mbox{SU(3)}_{\indR +\indL}$ generates eight Goldstone bosons.
They are not massless,
because the quark masses $m_u,m_d$ and $m_s$ break the symmetry, but since the
breaking is supposed to be small, these levels should remain lowest. Indeed,
the eight lightest bound
states, $\pi^+,\pi^0,\pi^-,$ $K^+,K^0,\bar{K}^0,K^-,\eta\,, $
do carry the required quantum numbers, both with respect to
spin/parity and to flavour.

As a further confirmation of the picture,
one may compare the mass splittings within the pseudoscalar octet with those of
the other multiplets. The mass {\em differences} are comparable: $M_\eta-M_\pi
\simeq
410\;\mbox{MeV},\;M_\Xi-M_N\simeq 380\;\mbox{MeV}$. The mass {\em ratios} of
the Goldstone bosons, however, deviate much more strongly from unity than those
of the other multiplets: while the various levels of the baryon octet
differ from their mean mass by less than
20 \%, the mass of the $\eta$ is four times as large as the mass
of the pion.
The above symmetry considerations neatly explain why this is
so. For ordinary multiplets, the eigenvalue of $H_0$ is
different from zero; the perturbation $H_{\mbox{\scriptsize sb}}$ only
generates
a correction, whose magnitude depends on the level in question, because
$H_{\mbox{\scriptsize sb}}$ breaks SU(3).
In the case of the Goldstone bosons, however,
the entire mass is due to the perturbation --- the pattern
of levels directly reveals the asymmetries of the operator
$H_{\mbox{\scriptsize sb}}$.
As
discussed above, $M_{\hspace{-0.07em}\pi^{\hspace{-0.07em}+}}$ is proportional
to \raisebox{0.25em}{$\sqrt{\rule{4em}{0em}}$}$\hspace{-4em}m_u+m_d\,$.
The same analysis applies to the currents $\overline{s}\gamma^\mu\gamma_5u$ and
$\overline{s}\gamma^\mu\gamma_5d$, which generate transitions from the vacuum
to the states $|K^+\rangle$ and $|K^0\rangle$.
Since the corresponding divergences are proportional
to $(m_u+m_s)$ and $(m_d+m_s)$, one now obtains
$M_{\hspace{-0.07em}K^{\hspace{-0.07em}+}}\!\propto\!$
\raisebox{0.25em}{$\sqrt{\rule{3.7em}{0em}}$}$\hspace{-3.7em}m_u+m_s\,$ and
$M_{\hspace{-0.07em}K^{\hspace{-0.07em}0}}\!\propto\!$
\raisebox{0.25em}{$\sqrt{\rule{3.5em}{0em}}$}$\hspace{-3.5em}m_d+m_s\,$.
The mass ratios of the Goldstone bosons strongly deviate from unity, because
$m_s$ happens to be large compared to $m_u$ and $m_d$.

The level shifts generated by the symmetry breaking may be analyzed by
treating the mass term in
eq.(\ref{pm2}) as a perturbation. To first order in the perturbation, the
result obeys the Gell-Mann-Okubo formula.
The calculation also applies to
the pseudoscalar
octet, where the unperturbed levels sit at $M\!=\!0$, provided the shifts in
the {\it square} of the mass are considered.
Indeed, $M_\pi^2,M_K^2$ and $M_\eta^2$ obey the formula remarkably well,
confirming that the mass pattern of
the pseudoscalar octet is perfectly consistent with the claim that SU(3) is a
decent approximate symmetry of the strong interaction.\footnote{The
experimental values of the decay
constants $F_\pi,F_K$, which represent the bound state wave functions
at the origin, also confirm the picture: The asymmetry seen there,
$F_K/F_\pi=1.22$ is quite
typical of the SU(3) breaking effects observed in other multiplets.}

The first order mass formulae for the pseudoscalar octet may also be used
to estimate the relative size of the three quark masses \cite{Weinberg77}.
The most remarkable feature of the resulting pattern is that the quark masses
strongly break isospin symmetry: $m_u$ and $m_d$ are quite
different \cite{GL75}. This may be verified as follows. Consider the
mass difference between
$K^0$ and $K^+$. If $m_u$ and $m_d$ where the same,
the splitting would exclusively be due to the electromagnetic
interaction. Since the main
contribution from this interaction is the self energy of the electric field
surrounding the $K^+$, this particle would have to be heavier than the $K^0$.
The observed splitting, $M_{K^0}-M_{K^+}= 4\;\mbox{MeV}$ is of opposite
sign. Hence the difference between $m_d$ and $m_u$ must make a
significant contribution, opposite to the electromagnetic one, $m_d>m_u$
(the same conclusion also follows from the mass
difference between neutron and proton). In first order perturbation theory,
the mass ratio $(M_{K^0}^2-M_{K^+}^2)/M_{\pi^+}^2$ is given by
the relative size of isospin breaking
in the quark masses, $r=(m_d-m_u)/(m_u+m_d)$.
Using the observed meson masses, this
gives $r\simeq0.20$. If the electromagnetic self energy is taken
into account, the result becomes even larger, because the two
contributions are of opposite sign: $r\simeq 0.29$ \cite{Weinberg77}.

The reason why, nevertheless, isopin is a nearly perfect symmetry of the strong
interaction is essentially the same
as for the case of SU(3) breaking, discussed above:
The relative magnitude of isospin breaking in the quark masses
does not represent an adequate estimate for the
magnitude of the isospin breaking effects occurring in the bound
states. What counts, instead, is the magnitude of the isospin breaking part of
the
Hamiltonian, $\Hbar_{\mbox{\scriptsize sb}}$, compared to the isospin symmetric
piece, $\Hbar_0$ (see eq.(\ref{iso1})). This is particularly evident
in the case of the nucleon, where the splitting
is of the order of $1\;\mbox{MeV}$, while the
isospin invariant part is responsible for the mean mass and is of order
$1\;\mbox{GeV}$. In algebraic terms, the matrix elements of
$\Hbar_{\mbox{\scriptsize sb}}$ are of order
$m_d-m_u$, while those of $\Hbar_0$ are determined by the scale
$\Lambda_{\QCD}$, so that the magnitude of isospin breaking is
determined by the ratio $(m_d-m_u)/\Lambda_{\QCD}$, rather than
$(m_d-m_u)/(m_u+m_d)$.

For the kaons, isospin breaking is enhanced, because these particles get their
mass from $m_s$, not from the scale of QCD: the
ratio $(M_{\hspace{-0.07em}K^{\hspace{-0.07em}0}}-
M_{\hspace{-0.07em}K^{\hspace{-0.07em}+}})/
(M_{\hspace{-0.07em}K^{\hspace{-0.07em}0}}+
M_{\hspace{-0.07em}K^{\hspace{-0.07em}+}})$ is of order $(m_d-m_u)/m_s$. One
might expect that the most important isospin breaking effects occur in the pion
multiplet, where the matrix elements of $\Hbar_0$ are suppressed even more
strongly. It so happens, however, that the strong breaking of SU(3) symmetry
seen in the pseudoscalar octet does not repeat itself here, because the matrix
elements of the perturbation,
$\langle\pi|\Hbar_{\mbox{\scriptsize sb}}|\pi\rangle$ are suppressed,
too: The mass splitting
$M_{\hspace{-0.07em}\pi^{\hspace{-0.07em}+}}-
M_{\hspace{-0.07em}\pi^{\hspace{-0.07em}0}}$ is of second order in the
perturbation, proportional to $(m_d-m_u)^2$.
Numerically, the effect is tiny, of order
$0.2\;\mbox{MeV}$; the observed mass difference is due almost entirely to the
electromagnetic interaction. The mathematical origin of this qualitative
difference between the two cases is that, in contrast to SU(3), the group
SU(2) does not have a $d$-symbol. For this reason, the pion mass is
shielded from isospin breaking, so that the range of the forces generated by
pion
exchange is nearly charge independent.
%%%%%%%%%%%%%%%%%%%%%%%%%%%%%%%%%%%%%%%%%%%%%%%%%%%%%%%%%%%%%%%%%%%
\section{Effective field theory}
\label{eff}
At low energies, the behaviour of scattering amplitudes or current matrix
elements can be described in terms of a {\it Taylor series expansion} in powers
of the momenta.
The electromagnetic form factor of the pion, e.g., may be
exanded in powers of the momentum transfer $t$.
In this case, the first two Taylor coefficients are related to the total charge
of the particle and to the mean square radius of the charge distribution,
respectively,
\be \label{taylor}
f_{\pi^+}(t) = 1 + \mbox{$\frac{1}{6}$} \langle r^2\rangle_{\pi^+}\, t +
O(t^2)\fs \end{equation}
Scattering lengths and effective ranges are analogous low energy
constants occurring in the Taylor series expansion of scattering amplitudes.

The occurrence of light particles gives rise to singularities in the low
energy domain, which limit the range of validity of the Taylor series
representation. The form factor $f_{\pi^+}(t)$, e.g., contains a branch cut
at $t=4 M_\pi^2$, such that the formula (\ref{taylor}) provides an adequate
representation only for $|t|\ll 4 M_\pi^2$. The problem becomes even more
acute if $m_u$ and $m_d$ are set equal to zero. The pion mass then
disappears, the branch cut sits at $t=0$ and the Taylor series does not
work at all. I first discuss the method used in the low energy analysis for
this extreme case, returning
to the physical situation with $m_u,m_d\neq 0$ below.

The reason why the spectrum of QCD with two massless quarks contains three
massless bound states is understood:
they are the Goldstone bosons of a
hidden symmetry. The symmetry, which
gives birth to these, at the same time also determines their low energy
properties. This makes it possible to explicitly work out
the poles and branch cuts generated by the exchange of Goldstone bosons.
The remaining singularities are
located comparatively far from the origin, the nearest one being due to
the $\rho$-meson. The result is a modified Taylor series expansion in powers
of the momenta, which works, despite the presence of massless particles.
In the case of the $\pi\pi$ scattering amplitude,
e.g., the radius of convergence of the modified series
is given by $s=M_\rho^2$, where $s$ is the square of the energy in the center
of mass system (the first few
terms of the series only yield a decent description of the amplitude if
$s$ is smaller than the radius of
convergence, say $s\!<\!\frac{1}{2}M_\rho^2\rightarrow \sqrt{s}\!<
540\;\mbox{MeV}$).

As pointed out by Weinberg \cite{Weinberg79}, the modified expansion
may explicitly be constructed by means of an effective field theory, which is
referred to as {\it chiral perturbation theory} and involves the following
ingredients: \\ (i) The quark and gluon fields of QCD are
replaced by a set of pion fields, describing the degrees of freedom of the
Goldstone
bosons. It is convenient to collect these in a
$2\!\times\!2$
matrix U$(x)\!\in\,$SU(2). \\
(ii) The Lagrangian of QCD is replaced by an
effective Lagrangian, which only involves the field U$(x)$, and its derivatives
\bdm {\cal L}_{\QCD}\;\;\longrightarrow \;\;{\cal L}_\eff(U,\partial
U,\partial^2U,\ldots)\fs\edm
(iii) The low energy expansion corresponds to an
expansion of the effective
Lagrangian, ordered according to the number of the derivatives of the field
$U(x)$.
Lorentz invariance only permits terms with an even number
of derivatives,
\bdm
{\cal L}_{\eff}= {\cal L}_{\eff}^{\,2} +
{\cal L}_{\eff}^{\,4} + {\cal L}_{\eff}^{\,6} +
\ldots
\edm

Chiral symmetry very strongly constrains the form of the terms occurring
in the series. In particular, it excludes momentum
independent interaction vertices:
Goldstone bosons can only interact if they carry momentum. This property
is essential for the consistency of the low energy analysis, which treats the
momenta as expansion parameters.
The leading contribution involves two derivatives,
\be \label{eff1}
{\cal L}_{\eff}^{\,2} = \mbox{$\frac{1}{4}$}F_\pi^2 \mbox{tr} \{
\partial_\mu U^+ \partial^\mu U \} \co
\ee
and is fully determined by the pion decay constant. At order $p^4$, the
symmetry permits two independent terms,\footnote{In the framework of the
effective theory, the anomalies of QCD
manifest themselves through an extra contribution,
the Wess-Zumino term, which is also of order $p^4$ and is proportional to the
number of colours.}
\be\label{eff3} {\cal L}_{\eff}^{\,4}=\mbox{$\frac{1}{4}$}l_1 (\mbox{tr} \{
\partial_\mu U^+ \partial^\mu U \})^2
+ \mbox{$\frac{1}{4}$}l_2\mbox{tr} \{
\partial_\mu U^+ \partial_\nu U \}\mbox{tr} \{
\partial^\mu U^+ \partial^\nu U \}\co\ee
etc. For most applications, the derivative expansion
is needed only to this order.

The most remarkable property of the method is that it does not
mutilate the theory under investigation:
The effective field theory framework is no more
than an efficient machinery, which
allows one to work out the modified Taylor series, referred to above.
If the effective Lagrangian includes all of the terms
permitted by the symmetry, the
effective theory is mathematically equivalent to QCD \cite{Weinberg79,found}.
It exclusively exploits the symmetry properties of QCD and involves an infinite
number of effective coupling constants,
$F_\pi,l_1,l_2,\ldots\;$, which represent the Taylor coefficients of the
modified expansion.

In QCD, the
symmetry, which controls the low energy properties of the Goldstone bosons, is
only an approximate one. The constraints imposed by the hidden,
approximate symmetry can still be worked out, at the price of
expanding the
quantities of physical interest in powers of the symmetry breaking parameters
$m_u$ and $m_d$. The low energy analysis then involves a combined
expansion,
which treats both, the momenta and the quark masses as small parameters.
The effective Lagrangian picks up additional terms, proportional to powers of
the quark mass matrix,
\bdm
m = \left(\mbox{\raisebox{0.4em}{$ m_u $}}\;\mbox{\raisebox{-0.4em}
{$m_d$}}\,
\right)
\edm
It is convenient to count $m$ like two powers of
momentum, such that the expansion of the effective Lagrangian still starts at
$O(p^2)$ and only contains even
terms. The leading contribution picks up a term linear in $m$,
\be\label{eff2}
{\cal L}_{\eff}^{\,2} = \mbox{$\frac{1}{4}$}F_\pi^2 \mbox{tr} \{
\partial_\mu U^+ \partial^\mu U \} +\mbox{$\frac{1}{2}$}F_\pi^2B\,\mbox{tr}
\{m(U+U^\dagger)\}\fs
\ee
Likewise, ${\cal L}_\eff^4$ receives additional contributions, involving two
further effective coupling constants, $l_3,l_4$, etc.

The expression (\ref{eff2}) represents a compact summary of the soft pion
theorems
established in the 1960's: The leading terms in the low energy expansion of the
scattering amplitudes and current matrix elements are given by the tree graphs
of this Lagrangian.
The coupling constant $B$, needed to account for the symmetry
breaking effects generated by the quark masses at leading order, represents the
coefficient of the linear term in the expansion of
the pion mass, $M_\pi^2=(m_u+m_d)B+O(m^2)$. According to section \ref{pm}, this
constant also determines the vacuum-to-pion matrix element of the
pseudoscalar density, $G_\pi =F_\pi B +O(m)$. Furthermore, the relation of
Gell-Mann, Oakes and Renner, $F_\pi^2M_\pi^2=-(m_u+m_d)\,\lvac\ubar u\rvac
+O(m^2)$,
which immediately follows from the
above expression for the effective Lagrangian, shows that the magnitude of
the quark condensate is also related to the value of $B$.

The effective field theory
represents an efficient and systematic framework, which allows one to work out
the corrections to the soft pion predictions, those arising from the
quark masses as well as those from the terms of higher order
in the momenta. The evaluation is based on a perturbative
expansion of the quantum fluctuations of the effective field. In addition to
the tree graphs relevant for the soft pion results, graphs containing vertices
from the higher order contributions ${\cal L}_\eff^4,{\cal L}_\eff^6\ldots$ and
loop graphs
contribute. The leading term of the effective Lagrangian describes
a nonrenormalizable theory, the "nonlinear $\sigma$-model". The
higher order terms in the derivative expansion, however, automatically contain
the relevant counter terms. The divergences occurring in the loop
graphs merely renormalize the effective coupling constants. The effective
theory is a perfectly renormalizable scheme, order by order in the low
energy expansion, so that, in principle, the result of the calculation does not
depend on who it is who did it.
%%%%%%%%%%%%%%%%%%%%%%%%%%%%%%%%%%%%%%%%%%%%%%%%%%%%%%%%%%%%%%%%%%%%
\section{Universality}
\label{uni}
The properties of the effective theory are governed by the hidden symmetry,
which is responsible for the occurrence of Goldstone bosons. In
particular, the form of the effective Lagrangian only depends on the symmetry
group G of the Hamiltonian and on the subgroup $\mbox{H}\subset \mbox{G}$,
under which the ground state is invariant. The Goldstone bosons live on the
difference between the two
groups, i.e., on the quotient G/H. The specific dynamical properties of the
underlying theory do not play any role. To discuss the consequences of this
observation,
I again assume that G is an exact symmetry.

In the case of QCD with two
massless quarks, $\mbox{G}=\mbox{SU(2)}_{\indR}\times\mbox{SU(2)}_{\indL}$ is
the group of
chiral isospin rotations, while $\mbox{H}=\mbox{SU(2)}$ is the ordinary isospin
group.
The Higgs model is another example of a theory with spontaneously broken
symmetry. It plays a crucial role in the Standard Model, where it describes
the generation of mass. The model involves a scalar field
$\vec{\phi}$ with four components. The Hamiltonian is invariant under
rotations of the vector $\vec{\phi}$, which form the group G = O(4). Since the
field picks up a vacuum expectation value, the
symmetry is spontaneously broken to the subgroup of those rotations,
which leave the vector $\lvac\vec{\phi}\rvac$ alone, H = O(3).
It so happens that these groups are the same as those above,
relevant for QCD.\footnote{The structure of the effective Lagrangian rigorously
follows from the
Ward identities for the Green functions of the currents, which also reveal the
occurrence of anomalies \cite{found}. The
form of the Ward identities is controlled by
the structure of G and H in the infinitesimal neighbourhood of the
neutral element. In this sense, the symmetry groups of the two models are the
same: O(4) and O(3) are {\it locally}
isomorphic to SU(2)$\times$SU(2) and
SU(2), respectively. }
The fact that the symmetries are the same implies that
the effective field theories are identical: (i) In either
case, there are three
Goldstone bosons, described by a matrix field $U(x)\in\mbox{SU(2)}$. (ii) The
form of the effective Lagrangian is precisely the same.
In particular, the expression
\bdm
{\cal L}_{\eff}^{\,2} = \mbox{$\frac{1}{4}$}F_\pi^2 \mbox{tr} \{
\partial_\mu U^+ \partial^\mu U \}
\edm
is valid in either case. At the level of the effective theory, the
only
difference between these two physically quite distinct models is that
the numerical values of the effective coupling constants are different.
In the case of QCD, the one occurring at leading order of the
derivative expansion is the pion decay constant, $F_\pi\simeq
93\,\mbox{MeV}$, while in the Higgs model, this coupling constant is larger
by more than three orders of magnitude, $F_\pi\simeq
250\;\mbox{GeV}$. At next-to-leading order, the effective coupling constants
are also different; in particular, in QCD, the anomaly coefficient is equal to
$\mbox{N}_c$, while in the Higgs model, it vanishes.

As an illustration, I compare the condensates of the two theories, which
play a role
analogous to the spontaneous magnetization $\langle\vec{M}\rangle$ of a
ferromagnet (or the staggered magnetization of an antiferromagnet).
At low temperatures, the magnetization singles out a direction --- the ground
state spontaneously breaks the symmetry
of the Hamiltonian with respect to rotations. As the system is heated, the
spontaneous magnetization decreases, because the thermal disorder acts against
the alignment of the spins. If the temperature is high enough, disorder
wins, the spontaneous magnetization disappears and rotational symmetry is
restored. The temperature at which this happens is the Curie temperature.
Quantities, which allow one to distinguish the ordered from the disordered
phase are called {\it order parameters}. The magnetization is the prototype of
such a parameter.

In QCD, the most important order parameter (the one of lowest dimension) is the
quark condensate. At nonzero temperatures, the condensate is given
by the thermal expectation
value \bdm \langle\ubar u\rangle_{\hspace{-0.05em}\mbox{\raisebox{-0.2em}
{\scriptsize $T$}}} =\frac{\mbox{Tr}\{\,\ubar u
\exp (-\,H/kT)\}}{ \mbox{Tr}\{\exp(-\,H/kT)\} }\fs
\edm
The condensate melts if the temperature is
increased. At a critical temperature, somewhere in the range
$140\,\mbox{MeV}\!<\!T_c\!<\!\mbox{180}\;\mbox{MeV}$, the quark condensate
disappears and chiral symmetry is restored. The same qualitative
behaviour also occurs in the Higgs model, where the expectation value
$\langle\,\vec{\phi}\,
\rangle_{\hspace{-0.05em}\mbox{\raisebox{-0.2em}
{\scriptsize $T$}}}$ of the scalar field represents the most prominent order
parameter.

At low temperatures, the thermal trace is dominated by
states of low energy. Massless particles generate contributions which are
proportional to powers of the temperature, while massive ones like the
$\rho$-meson are suppressed by the corresponding Boltzmann factor,
$\exp(-M_\rho/kT)$. In the case of a spontaneously broken symmetry,
the massless particles are the Goldstone bosons and their contributions may be
worked out by means of effective field theory. For the quark condensate, the
calculation has been done \cite{Gerber}, up to and including terms of order
$T^6$: \bdm \langle\ubar u\rangle_{\hspace{-0.05em}\mbox{\raisebox{-0.1em}
{\scriptsize $T$}}} =
\lvac\ubar u\rvac\!
\left\{1\,-\,\frac{T^2}
{8F_\pi^2 }
\,-\,\frac{T^4}{384F_\pi^4}
\,-\,\frac{T^6}{288F_\pi^6}
\, \ln(T_1/T)
\,+\,O(T^8)\right\}\fs\edm
The formula is exact --- for massless quarks, the temperature scale relevant
at low $T$ is the pion decay constant. The additional logarithmic scale $T_1$
occurring at order $T^6$ is determined by the effective coupling constants
$l_1,l_2$, which enter the expression (\ref{eff3}) for the effective Lagrangian
of order $p^4$. Since these are known from the phenomenology of $\pi\pi$
scattering, the value of $T_1$ is also known:
$T_1=470\pm110\;\mbox{MeV}$.

Now comes the point I wish to make. The effective Lagrangians
relevant for QCD and for the Higgs model are the same. Since the
operators of which we are considering the expectation values also transform in
the
same manner, their low temperature expansions are identical. The above formula
thus holds, without any change whatsoever, also for the Higgs condensate,
\bdm \langle\,\vec{\phi}\,\rangle_{\hspace{-0.1em}\mbox{\raisebox{-0.2em}
{\scriptsize $T$}}} =
\lvac\vec{\phi}\rvac\!
\left\{1\,-\,\frac{T^2}
{8F_\pi^2 }
\,-\,\frac{T^4}{384F_\pi^4}
\,-\,\frac{T^6}{288F_\pi^6}
\, \ln(T_1/T)
\,+\,O(T^8)\right\}\fs\edm
In fact, the universal term of order $T^2$ was discovered in the framework of
this model, in connection with work on the electroweak phase transition
\cite{Binetruy}.

The effective Lagrangian of a Heisenberg
antiferromagnet is also of the same structure,\footnote{Since the ground state
of a magnet fails to be Lorentz invariant, the derivative expansion of
the effective Lagrangian contains additional contributions. For a
cubic lattice, however, the leading term is of the same form as in
relativistically invariant theories, except that the
velocity of light is to be replaced by the velocity of propagation for magnons
of long wavelength. The low energy properties of a
ferromagnet, on the other hand,
are quite different. The corresponding effective Lagrangian is
dominated by a topological term, related to the fact that the generators
of the symmetry acquire nonzero expectation values in the ground state
\cite{Randjbar}.}
so that the above formula even holds for the staggered magnetization, except
for one modification:
the Clebsch-Gordan coefficients, which accompany the various powers of $T$ are
different, because the symmetry groups differ: The Hamiltonian now is
invariant under ordinary rotations, G = O(3), while the
ground state
spontaneously breaks the symmetry to the subgroup H = O(2) of the rotations
around the direction singled out by
the magnetization.

These examples illustrate the physical nature of effective theories: At long
wavelength, the microscopic structure does not play any role. The behaviour
only depends on those degrees of freedom, which require little
excitation
energy. The hidden symmetry, which is responsible for the absence of an
energy gap and for the occurrence of Goldstone bosons, at the same time also
determines their low energy properties. For this reason, the form of
the effective Lagrangian is controlled
by the symmetries of the system and is, therefore, universal.
The microscopic structure of the underlying theory exclusively manifests itself
in the numerical values of the effective coupling constants.
The temperature expansion also clearly exhibits the limitations of
the method. The truncated series can be trusted only at low temperatures,
where the first term represents the dominant contribution. According to the
above formula, the quark condensate drops to about half of the vacuum
expectation value when the temperature reaches
$160\;\mbox{MeV}$ --- the formula does not make much sense beyond this
point. In particular, the behaviour of
the quark condensate in the vicinity of the chiral phase transition is
beyond the reach of the effective theory discussed here.
%%%%%%%%%%%%%%%%%%%%%%%%%%%%%%%%%%%%%%%%%%%%%%%%%%%%%%%%%%%%%%%%%%%%%%%%%%%
\section{Experimental aspects}
\label{exp}
The DAFNE Handbook \cite{Dafne} provides an excellent overview over many of the
processes, where new data will contribute to make progress in
understanding the low energy structure of QCD. I only add a few comments.

One of the issues, about which very little is
known experimentally, is the explicit breaking of chiral and isospin
symmetry, generated by $m_u$ and $m_d$.
Because the group SU(2)$_{\indR}\times$SU(2)$_{\indL}$ represents
an almost exact symmetry of the strong interaction,
the symmetry breaking part of the Hamiltonian only generates very
small effects.
An excellent place to check the theoretical ideas about the implications
of symmetry breaking is $\pi\pi$ scattering. As shown by
Weinberg \cite{Weinberg66},
nearly 30 years ago, chiral symmetry leads to parameter free soft pion
predictions
for the corresponding $S$-wave scattering lengths $a_0,a_2$. There is a
beautiful
proposal \cite{Nemenov} to accurately measure the combination $a_0-a_2$, by
producing $\pi^+\pi^-$ atoms and measuring the rate of their decay into
$\pi^0\pi^0$. The corrections to the soft pion results have been worked out
\cite{GL83}, so that a very accurate prediction is available for test.
The $S$-wave scattering lengths are closely related to the $\sigma$-term matrix
element $\sigma_{\hspace{-0.05em}\pi\hspace{-0.05em}
\pi}=\langle\pi|m_u\ubar
u+m_d\dbar d|\pi\rangle$ and
are also proportional
to $m_u+m_d$. The quantity $a_0-a_2$ thus represents a direct measure of the
asymmetries produced by the quark
masses. The experiment, in particular, would provide a sensitive test of the
standard hypothesis, according to which the expansion of the pion mass in
powers of the quark masses,
\bdm M_\pi^2=M^2\left\{1-
\frac{M^2}{32\pi^2 F_\pi^2}\,\lbar_3+O(M^4)\right\}\;\;,\hspace{2em}
M^2\equiv(m_u+m_d)B\co\edm
is dominated by the first term. In the standard
picture, the contribution of order $(m_u+m_d)^2$, which is
proportional to the effective coupling constant $\lbar_3$, amounts to a
small
correction of order 2\%; the corresponding contribution to $a_0-a_2$ is three
times smaller. As pointed out by Knecht et al. \cite{Knecht}, the
arguments which underly this estimate are theoretical: There is no direct
experimental evidence,
which would rule out an entirely different picture. A
number like $\lbar_3=-100$, e.g., would
increase the result for $a_0-a_2$ by about 25\% and bring it into agreement
with the central value of the currently available data. Conversely, if this
value should be confirmed within narrow error bars,
one would have to conclude that the "correction" in
the expansion of $M_\pi^2$ is almost as large as the leading
term.
Needless to
say that this would give rise to a major earthquake in the current
understanding
of QCD. The quark mass pattern discussed above is based on the standard
picture, where it is assumed
that the Gell-Mann-Oakes-Renner relation is not
ruined by higher order terms. This is the only way I know of to
understand the success of the Gell-Mann-Okubo formula for the pseudoscalar
octet --- if the symmetry breaking observed in $\pi\pi$
scattering should disagree with the theoretical predictions, the
standard picture would require thorough revision, even at the qualitative
level. Only few of us expect this to be the outcome of the investigation, but
the earmark of an important experiment is the product of the likelihood for
a discovery with the physical significance thereof, the likelihood as
such may be quite small.

The analogous issue in pion-nucleon scattering is a dynosaur. It is
notoriously difficult to accurately measure
$\sigma_{\hspace{-0.05em}\pi\hspace{-0.1em}
{\scriptscriptstyle N}}$.
At the present
time, the experimental uncertainties in this quantity amount to
about
20\%, comparable to those in the $\pi\pi$ $S$-wave scattering lengths. There
are beautiful new data on the related $\pi N$ scattering lengths,
based on bound states of $\pi^-p$ and $\pi^-d$ \cite{Leisi},
analogous to the $\pi^+\pi^-$ atoms of
the proposal mentioned above. These data attain
a precision, where even isospin
breaking effects due to $m_d-m_u$ can be measured,
provided the theoretical
results \cite{threebody}, used to
the express the pion-deuteron scattering lengths in terms of
those
of proton and neutron, can be trusted at the accuracy needed here.
The new data should
give ample incentive for a careful reanalysis of the three-body problem,
which arises if a pion of zero momentum encounters a deuteron.
Evidently, the experimental
discrepancies in low energy $\pi N$ scattering should be resolved.
For a measurement of small quantitities like
$\sigma_{\hspace{-0.05em}\pi\hspace{-0.1em}
{\scriptscriptstyle N}}$, the
dominating contribution from the Born
term, i.e., the value of the coupling constant
$g_{\pi\hspace{-0.1em}
{\scriptscriptstyle N}}$, needs to be known
to very high precision.

On the
theoretical side, considerable progress in the chiral perturbation
theory of the $\pi N$ interaction is being made. The predictions are
weaker
here, because, in the hidden symmetry game, the nucleons are only
spectators, not actors like the Goldstone bosons.
Accordingly, the number of effective coupling constants, which need
to be taken from phenomenology, is larger. In the case of the $\sigma$-term,
e.g., the symmetry implies that the matrix element $\langle\pi|\,\sbar
s|\pi\rangle$ vanishes if $m_u,m_d$ are sent to zero, while this is not the
case for the corresponding nucleon matrix element.
Also, the $\pi\pi$ scattering matrix
elements are shielded from the perturbations generated by $m_d-m_u$, but the
$\pi N$ scattering matrix elements are not ---
{\it small}
quantities like the $\sigma$-term or the isospin even $S$-wave scattering
length
may pick up {\it comparatively} large charge asymmetries \cite{Weinberg77}.
The fact that the excitation
energy of the $\Delta$ is relatively small does not really present a problem;
unless one attempts to use the effective theory in the
vicinity of the resonance or beyond, the
corresponding singularity may be expanded in the standard fashion, absorbing
the Taylor coefficients in the relevant low energy constants. The expansion
of the $\pi N$ scattering amplitude in powers of the momenta, however,
contains odd as well as even powers ---
one needs to carry the expansion beyond the first two terms to
achieve the same precision as the one available for $\pi\pi$ scattering
\cite{Matsinos}. Work on this problem is of interest, in particular, in
connection with the ongoing experiments on pion photo-
and electroproduction, whose significance as probes of the low energy
structure is becoming increasingly evident and which were discussed in detail
at this workshop.

Another topic, where the experimental situation needs to be clarified, is
$\eta$
decay. It is important to resolve the discrepancy between the older data, based
on the Primakoff effect and the more recent ones, from
photon-photon-collisions. The rate of the decay into three pions measures the
ratio
$(m_d^2-m_u^2)/m_s^2$ of quark masses \cite{eta}. Also, the
available information on the Dalitz plot distribution of
the $\pi^+\pi^-\pi^0$ final state and on the ratio
$\Gamma_{\eta\rightarrow 3\pi^0}/
\Gamma_{\eta\rightarrow \pi^0\pi^+\pi^-}$
leaves to be desired. Incidentally, the world average of the partly
inconsistent
data on these quantities is not in satisfactory agreement with the theoretical
predictions.

There are many other items of interest, which are by no means less
interesting --- processes generated by
the Wess-Zumino term, to only name one category --- but I stop here, thanking
Aron Bernstein, Barry Holstein and their coworkers for a very informative
meeting.
%%%%%%%%%%%%%%%%%%%%%%%%%%%%%%%%%%%%%%%%%%%%%%%%%%%%%%%%%%%%%%%%%%%%%%

\end{document}